\documentstyle[12pt,epsfig]{article}

\catcode`\@=11
\def\@citex[#1]#2{%
\if@filesw \immediate \write \@auxout {\string \citation {#2}}\fi
\@tempcntb\m@ne \let\@h@ld\relax \def\@citea{}%
\@cite{%
  \@for \@citeb:=#2\do {%
    \@ifundefined {b@\@citeb}%
      {\@h@ld\@citea\@tempcntb\m@ne{\bf ?}%
      \@warning {Citation `\@citeb ' on page \thepage \space undefined}}%
      {\@tempcnta\@tempcntb \advance\@tempcnta\@ne%
      \@tempcntb\number\csname b@\@citeb \endcsname \relax%
      \ifnum\@tempcnta=\@tempcntb 
        \ifx\@h@ld\relax%
          \edef \@h@ld{\@citea\csname b@\@citeb\endcsname}%
        \else%
          \edef\@h@ld{\ifmmode{-}\else--\fi\csname b@\@citeb\endcsname}%
        \fi%
      \else
        \@h@ld\@citea\csname b@\@citeb \endcsname%
        \let\@h@ld\relax%
      \fi}%
    \def\@citea{,\penalty\@highpenalty\,}%
  }\@h@ld
}{#1}}

\def\@citeb#1#2{{[#1]\if@tempswa , #2\fi}}
\def\@citeu#1#2{{$^{#1}$\if@tempswa , #2\fi }}
\def\@citep#1#2{{#1\if@tempswa , #2\fi}}

\def\bcites{         
        \catcode`\@=11
        \let\@cite=\@citeb
        \catcode`\@=12
}

\def\upcites{         
        \catcode`\@=11
        \let\@cite=\@citeu
        \catcode`\@=12
}

\def\plaincites{      
        \catcode`\@=11
        \let\@cite=\@citep
        \catcode`\@=12
}

\newcount\hour
\newcount\minute
\newtoks\amorpm
\hour=\time\divide\hour by 60
\minute=\time{\multiply\hour by 60 \global\advance\minute by-\hour}
\edef\standardtime{{\ifnum\hour<12 \global\amorpm={am}%
        \else\global\amorpm={pm}\advance\hour by-12 \fi
        \ifnum\hour=0 \hour=12 \fi
        \number\hour:\ifnum\minute<10 0\fi\number\minute\the\amorpm}}
\edef\militarytime{\number\hour:\ifnum\minute<10 0\fi\number\minute}

\def\draftlabel#1{{\@bsphack\if@filesw {\let\thepage\relax
   \xdef\@gtempa{\write\@auxout{\string
      \newlabel{#1}{{\@currentlabel}{\thepage}}}}}\@gtempa
   \if@nobreak \ifvmode\nobreak\fi\fi\fi\@esphack}
        \gdef\@eqnlabel{#1}}
\def\@eqnlabel{}
\def\@vacuum{}
\def\marginnote#1{}
\def\draftmarginnote#1{\marginpar{\raggedright\scriptsize\tt#1}}
\def\draft{
        \pagestyle{plain}
        \overfullrule=2pt
        \oddsidemargin -.5truein
        \def\@oddhead{\sl \phantom{\today\quad\militarytime} \hfil
        \smash{\Large\sl DRAFT} \hfil \today\quad\militarytime}
        \let\@evenhead\@oddhead
        \let\label=\draftlabel
        \let\marginnote=\draftmarginnote
        \def\ps@empty{\let\@mkboth\@gobbletwo
        \def\@oddfoot{\hfil \smash{\Large\sl DRAFT} \hfil}
        \let\@evenfoot\@oddhead}
        \def\@eqnnum{(\theequation)\rlap{\kern\marginparsep\tt\@eqnlabel}%
        \global\let\@eqnlabel\@vacuum}  }

\def\blackfonts{
        \font\blackboard=msbm10 scaled\magstep1
        \font\blackboards=msbm8
        \font\blackboardss=msbm6
}

\def\nblack{            
        \def\ZZ{{Z \n{10} Z}}
        \def\NN{{N \n{14} N}}
        \def\CC{{C \n{11} C}}
        \def\RR{{R \n{11} R}}
        \def\QQ{{Q \n{12} Q}}
        \def\PP{{P \n{11} P}}
}

\def\prep{         
        \catcode`\@=11
        \input art10.sty
        \catcode`\@=12
        
        \let\small\null
        \def\blackfonts{
                \font\blackboard=msbm10
                \font\blackboards=msbm7
                \font\blackboardss=msbm5
        }
        \let\sl\it
        \twocolumn
        \sloppy
        \voffset=-2.54truecm
        \hoffset=-2.54truecm
        \flushbottom
        \parindent 1em
        \leftmargini 2em
        \leftmarginv .5em
        \leftmarginvi .5em
        \marginparwidth 48pt
        \marginparsep 10pt
        \setlength{\columnsep}{2truecm}
        \setlength{\textwidth}{25.4truecm}
        \setlength{\textheight}{17truecm}
        \baselineskip=16pt
        \oddsidemargin .18truein
        \evensidemargin .17truein
}

\def\eqalign#1{\null\,\vcenter{\openup\jot\m@th
  \ialign{\strut\hfil$\displaystyle{##}$&$\displaystyle{{}##}$\hfil
      \crcr#1\crcr}}\,}
\def\eqalignno#1{\displ@y \tabskip\centering
  \halign to\displaywidth{\hfil$\@lign\displaystyle{##}$\tabskip\z@skip
    &$\@lign\displaystyle{{}##}$\hfil\tabskip\centering
    &\llap{$\@lign##$}\tabskip\z@skip\crcr
    #1\crcr}}

\def\section{\@startsection {section}{1}{\z@}{3.ex plus 1ex minus
 .2ex}{2.ex plus .2ex}{\large\bf}}
\def\subsection{\@startsection{subsection}{2}{\z@}{2.75ex plus 1ex minus
 .2ex}{1.5ex plus .2ex}{\bf}}        

\def\appendix{{\newpage\section*{Appendix}}\let\appendix\section%
        {\setcounter{section}{0}
        \gdef\thesection{\Alph{section}}}\section}

\def\abstract{\if@twocolumn
\section*{Abstract}
\else 
\begin{center}
{\bf Abstract\vspace{-.5em}\vspace{0pt}}
\end{center}
\quotation
\fi}

\catcode`\@=12

\newcommand{\beq}{\begin{equation}}
\newcommand{\eeq}{\end{equation}}
\newcommand{\beqa}{\begin{eqnarray}}
\newcommand{\eeqa}{\end{eqnarray}}

\def\noj#1,#2,{{\bf #1} (19#2)\ }
\def\jou#1,#2,#3,{{\sl #1\/ }{\bf #2} (19#3)\ }
\def\ann#1,#2,{{\sl Ann.\ Physics\/ }{\bf #1} (19#2)\ }
\def\cmp#1,#2,{{\sl Comm.\ Math.\ Phys.\/ }{\bf #1} (19#2)\ }
\def\ma#1,#2,{{\sl Math.\ Ann.\/ }{\bf #1} (19#2)\ }
\def\ng#1,#2,{{\sl Nagoya.\ Math.\ J.\/ }{\bf #1} (19#2)\ }
\def\jd#1,#2,{{\sl J.\ Diff.\ Geom.\/ }{\bf #1} (19#2)\ }
\def\invm#1,#2,{{\sl Invent.\ Math.\/ }{\bf #1} (19#2)\ }
\def\cq#1,#2,{{\sl Class.\ Quantum Grav.\/ }{\bf #1} (19#2)\ }
\def\cqg#1,#2,{{\sl Class.\ Quantum Grav.\/ }{\bf #1} (19#2)\ }
\def\ijmp#1,#2,{{\sl Int.\ J.\ Mod.\ Phys.\/ }{\bf A#1} (19#2)\ }
\def\jmphy#1,#2,{{\sl J.\ Geom.\ Phys.\/ }{\bf #1} (19#2)\ }
\def\jams#1,#2,{{\sl J.\ Amer.\ Math.\ Soc.\/ }{\bf #1} (19#2)\ }
\def\grg#1,#2,{{\sl Gen.\ Rel.\ Grav.\/ }{\bf #1} (19#2)\ }
\def\mpl#1,#2,{{\sl Mod.\ Phys.\ Lett.\/ }{\bf A#1} (19#2)\ }
\def\nc#1,#2,{{\sl Nuovo Cim.\/ }{\bf #1} (19#2)\ }
\def\np#1,#2,{{\sl Nucl.\ Phys.\/ }{\bf B#1} (19#2)\ }
\def\pl#1,#2,{{\sl Phys.\ Lett.\/ }{\bf #1B} (19#2)\ }
\def\pla#1,#2,{{\sl Phys.\ Lett.\/ }{\bf #1A} (19#2)\ }
\def\pr#1,#2,{{\sl Phys.\ Rev.\/ }{\bf #1} (19#2)\ }
\def\prd#1,#2,{{\sl Phys.\ Rev.\/ }{\bf D#1} (19#2)\ }
\def\prl#1,#2,{{\sl Phys.\ Rev.\ Lett.\/ }{\bf #1} (19#2)\ }
\def\prp#1,#2,{{\sl Phys.\ Rept.\/ }{\bf #1C} (19#2)\ }
\def\ptp#1,#2,{{\sl Prog.\ Theor.\ Phys.\/ }{\bf #1} (19#2)\ }
\def\ptpsup#1,#2,{{\sl Prog.\ Theor.\ Phys.\/ Suppl.\/ }{\bf #1} (19#2)\ }
\def\rmp#1,#2,{{\sl Rev.\ Mod.\ Phys.\/ }{\bf #1} (19#2)\ }
\def\yadfiz#1,#2,#3[#4,#5]{{\sl Yad.\ Fiz.\/ }{\bf #1} (19#2) #3%
\ [{\sl Sov.\ J.\ Nucl.\ Phys.\/ }{\bf #4} (19#2) #5]}
\def\zh#1,#2,#3[#4,#5]{{\sl Zh.\ Exp.\ Theor.\ Fiz.\/ }{\bf #1} (19#2) #3%
\ [{\sl Sov.\ Phys.\ JETP\/ }{\bf #4} (19#2) #5]}

\def\beq{\begin{equation}}
\def\eeq{\end{equation}}
\def\beqar{\begin{eqnarray}}
\def\eeqar{\end{eqnarray}}

\newcommand{\be}{\begin{equation}}
\newcommand{\ee}{\end{equation}}
\newcommand{\bea}{\begin{eqnarray}}
\newcommand{\eea}{\end{eqnarray}}

\def\nfrac#1#2{{\displaystyle{\vphantom1\smash{\lower.5ex\hbox{\small$#1$}}%
        \over\vphantom1\smash{\raise.25ex\hbox{\small$#2$}}}}}

\def\n#1{\mskip-#1mu}

\def\lae{\mathrel{\mathop{\smash{\lower .5 ex \hbox{$\stackrel<\sim$}}}}}
\def\lae{\mathrel{\mathop{\smash{\lower .5 ex \hbox{$\stackrel>\sim$}}}}}

\def\tr{{\rm Tr}}
\def\l:{\mathopen{:}\,}
\def\r:{\,\mathclose{:}}

\catcode`\@=11
\def\theequation{\arabic{equation}}
\catcode`\@=12

\nblack
\bcites

\nblack

\catcode`\@=11
\def\theequation{\thesection.\arabic{equation}}
\@addtoreset{equation}{section}
\@addtoreset{footnote}{section}
\@addtoreset{footnote}{subsection}
\catcode`\@=12

\newcommand{\beqn}{\begin{equation}}
\newcommand{\eeqn}{\end{equation}}
\newcommand{\beqnarray}{\begin{eqnarray}}
\newcommand{\eeqnarray}{\end{eqnarray}}
\newcommand {\bear} [1] {\begin {array} {#1}}
\newcommand {\ear} {\end {array}}

\newcommand {\beqarn} {\begin{eqnarray*}}
\newcommand {\eeqarn} {\end{eqnarray*}}

\newcommand{\ccdot}{\! \cdot \!}

\begin{document}

\begin{titlepage}

\begin{center}
\today
\hfill LBNL-41931, UCB-PTH-98/32\\
\hfill                  hep-th/9806104

\vskip 1.5 cm
{\large \bf Six-Dimensional Supergravity on $S^3\times AdS_3$\\
and 2d Conformal Field Theory}\\
\vskip 1 cm 
{Jan de Boer}\\
\vskip 0.5cm
{\sl Department of Physics,
University of California at Berkeley\\
366 Le\thinspace Conte Hall, Berkeley, CA 94720-7300, U.S.A.\\
and\\
Theoretical Physics Group, Mail Stop 50A--5101\\
Ernest Orlando Lawrence Berkeley National Laboratory\\
Berkeley, CA 94720, U.S.A.\\}

\end{center}

\vskip 0.5 cm

\begin{abstract}

In this paper we study the relation between
six-dimensional supergravity compactified
on $S^3 \times AdS_3$ and certain two-dimensional
conformal field theories. We compute the Kaluza-Klein
spectrum of supergravity using representation theory;
these methods are quite general and can also be applied
to other compactifications involving anti-de Sitter spaces.
A detailed comparison between the spectrum of the 
two-dimensional conformal field theory and supergravity is
made, and we find complete agreement. This applies even
at the level of certain non-chiral primaries, and we propose a
resolution to the puzzle of the missing states recently
raised by Vafa. As a further illustration of the method
the Kaluza-Klein spectra of F-theory on 
$M^6\times S^3 \times AdS_3$ and of M-theory on
$M^6\times S^2 \times AdS_3$ are computed, with $M^6$ 
some Calabi-Yau manifold.

\end{abstract}

\end{titlepage}

\section{Introduction}

One of the most interesting examples of the AdS$\leftrightarrow$CFT 
conjecture proposed in \cite{mal} and refined in \cite{wit,gkp} is
the duality between type IIB string theory on $M^4\times S^3 \times AdS_3$
and certain two-dimensional conformal field theories. 
Two-dimensional conformal field theories are very well understood,
enabling us to test the conjecture in more detail than in other dimensions.
Furthermore, by S-dualizing the type IIB background we obtain a string
theory with only NS-NS fields turned on, which looks like the product
of a $K3$ conformal field theory and an $SU(2)$ and $Sl(2,{\bf R})$ 
WZW theory. Thus we may hope to better understand the AdS side
of the conjecture as well, although after the S-duality the type IIB
string theory is strongly coupled and it is not clear to what extent
we can trust naive conformal field theory considerations. 

The main example of the conjecture arises when we consider a system
of parallel D1 and D5 branes, where the D5 branes are wrapped on some
four manifold $M^4$ which can be either $T^4$ or $K3$. 
The solitonic description of this brane configuration yields a
five-dimensional black hole. The relation between the D1-D5 brane
system and the five-dimensional black hole has been examined in
great detail. In particular, the D1-D5 brane system correctly
accounts for the entropy of the black hole \cite{vast} and
various emission and absorption probabilities and 
greybody factors
\cite{dama,gukl,mast2,homast}.
The D1-D5 brane system is described by a certain 1+1 dimensional
gauge theory. 
Related gauge theories 
appear in the study of the
M5 brane in M(atrix) theory, in the M(atrix) 
description of M-theory on $T^5$, and in the description of
little string theories, and were studied in
\cite{dvv,dvv2,sei1,dps,bedo,abkss,wit3,dise,dou,hawa,malphd}. For all these application it
is important to know to which conformal field theory the gauge
theory flows in the infrared. In \cite{wit3} it was argued that
in the infrared the Coulomb and Higgs branches of the gauge
theory decouple. For the case where the D5 branes are wrapped
over $M^4$, the conformal field theory is a deformation of the
$N=(4,4)$ sigma model with target space $(M^4)^N/S_N$ \cite{va1,vast}.
Going to the infrared for the gauge theory is the same as
going to the near horizon region of the five-dimensional black
hole \cite{mal}, whose geometry is 
$M^4\times S^3 \times AdS_3$\footnote{There has been a lot of
work on the relation of $AdS_3$ to black holes, entropy, and
brane configurations; see 
\cite{btz,hyun,skend1,lee,hosom,skend2,strom,lars,alw,kalo,ioza,brun,sachs,johns,teo,wzw2,carlip9} 
for an incomplete list
of references.}.
Thus, type IIB string theory in this background should be
dual to a 2d conformal field theory associated to $(M^4)^N/S_N$.
In fact, most of the D1-D5 brane calculations rely only on
the IR conformal field theory description of the gauge theory on
the branes, and support this duality.
Further evidence and results were
recently obtained in \cite{mastr,mart}.

One of the goals of this paper is to compare in detail the 
spectrum of the conformal field theory and 6d (2,0)
supergravity\footnote{We
can restrict attention to six-dimensional supergravity because
the radius of $M^4$ is much smaller than that of $S^3$ or
$AdS_3$.}
on $S^3 \times AdS_3$.  This
is of particular interest in view of the recent paper \cite{vafap}
in which it is argued that certain states in the conformal
field theory are absent in supergravity. We will see that
the Kaluza-Klein spectrum of supergravity can be completely
determined using only representation theory, and in particular
we can find the masses and spins of all KK-fields on $AdS_3$,
without the need to examine the equations of motion of supergravity.
A careful comparison between the multi-particle KK spectrum of
supergravity and the spectrum of the conformal field theory 
reveals a complete agreement and automatically suggests
a resolution to the puzzle of \cite{vafap}.

Another goal of this paper is to examine to what extent 
six-dimensional supergravity on $S^3 \times AdS_3$ is 
always dual to some conformal
field theory on the boundary of $AdS_3$, as arguments based
on holography suggest \cite{wisu}. To do this, we compute the
KK spectrum of arbitrary 6d supergravities on $S^3 \times AdS^3$.
This includes cases that do not obviously correspond to the near-horizon
geometry of some brane configuration.
For all six-dimensional supergravities with $N>1$ supersymmetry, the
KK spectrum allows for a straightforward interpretation in
terms of some conformal field theory with target space $(M^4)^N/S_N$.
The case with $N=1$ supersymmetry is more mysterious. 
For six-dimensional supergravities
obtained from F-theory on a Calabi-Yau three-manifold $M^6$,
the KK spectrum is organized naturally in terms of the Betti numbers
of the mirror Calabi-Yau $\tilde{M}^6$. The precise meaning of this
remains to be understood. However, in view of the duality between
F-theory on $M^6$ and type I on $K3$, it is natural to identify
F-theory on $M^6 \times S^3 \times AdS_3$ with the near horizon
geometry of a D1-D5 brane system in type I. The $(0,4)$ theory
on the boundary of $AdS_3$ would then be the IR description of
the gauge theory on the D1 branes.

With the same goal in mind, 
we will also briefly consider five-dimensional supergravity on
$S^2 \times AdS_3$; the techniques in this paper are however
rather general and can be extended to other compactifications 
involving some anti-de Sitter space $AdS_p$. 

The outline of this paper is as follows.

In section~2 we will briefly review the D1-D5 system wrapped on
$M^4$, and its near horizon geometry.

In section~3, we discuss some generic features of six-dimensional
supergravity on $S^3\times AdS_3$. We will explain the relation 
between the global symmetries of supergravity and the chiral
algebra of the conformal field theory living at the boundary of $AdS_3$.
We will also derive the relation between the conformal weights of
fields at the boundary, and the masses and spins of the fields in the
bulk. 

In section~4, we explain the general idea how to use representation
theory to determine the KK spectrum, and 
use the oscillator method to
determine the short representations
of the AdS supergroup $SU(1,1|2)$ relevant for this problem.

In section~5 we discuss the spectrum of the conformal field theory
at the boundary, by considering the cohomology and elliptic genus
of a symmetric product. 

In section~6, we compute the KK spectrum for the chiral $N=(2,0)$ 
$d=6$ supergravity which describes type IIB supergravity compactified
on $K3$. The multiparticle spectrum of states
in supergravity contains all chiral primaries of the conformal
field theory, but also various non-chiral primaries. 
We check in an example that these non-chiral primaries
occur with the right multiplicity as predicted by the 
elliptic genus, and propose that they constitute the missing
states of \cite{vafap}.

In section~7, other six-dimensional supergravities 
compactified on $S^3 \times AdS_3$ are considered, and a preliminary
discussion of five-dimensional supergravity on $S^2 \times AdS^3$ is
given.

Finally, we give some conclusions in section~8.

Several of the results in this paper were announced in \cite{talk}.
While this work was nearing completion, a paper \cite{larsen} appeared that
also studies KK spectra using representation theory.

\section{The D1-D5 Brane System}

The metric for the extremal system of $Q_1$ D1 branes and $Q_5$ D5 branes
wrapped on K3 is given by \cite{homast,mastr}
\bea
\frac{ds^2}{\alpha'} & = & \frac{U^2}{\ell^2}
(-dt^2 + (dx^5)^2) + \frac{\ell^2}{U^2}dU^2
\nonumber \\
& & + \ell^2 d\Omega_3^2 \\
& & + \sqrt{\frac{Q_1}{v Q_5}} ds^2_{K3}
\eea
where $\ell^2 = g_6\sqrt{N}$, $N=Q_1 Q_5$, and $ds^2_{K3}$ is the metric
on $K3$ with volume $16\pi^4 v \alpha'^2$. The dilaton is given by
$\exp(-2 \phi) =  Q_5/g_6^2 Q_1$, with $g_6$ the six-dimensional
and $g=g_6 \sqrt{v}$ the ten-dimensional 
 string
coupling constant. The RR three-form field strength $H$ satisfies
$\int_{S^3} H = \int_{S^3} \ast_6 H =  4 \pi^2 \alpha' Q_5$. 

The limit in which we trust the supergravity approximation is the
one where we keep $g_6 Q_i$ large and fixed, and send $g_6$
to zero. Because ${\rm vol}(K3)/\alpha'^2 \ell^4\sim 1/(g_6 Q_5)^2$
is very small for these values of the parameters, only the massless
modes of supergravity on $K3$ will be important, giving rise to 
a six-dimensional supergravity theory with $(2,0)$ supersymmetry
and 21 tensor multiplets, compactified on $S^3 \times AdS_3$.

Several of the moduli of K3 in the near horizon limit are independent
of their values of infinity in the original type IIB setup. The
values of these fixed scalars, such as the volume of K3, are
a function of the charges only and determined by minimizing a suitable
central charge as in \cite{feka,andafe}. Since the central charge
is linear in $Q_1$ and $Q_5$, rescaling both by a fixed factor will
rescale the central charge by the same factor and therefore not
affect the moduli of the K3. Hence, the K3 will generically be
a smooth K3, and the sizes of the 
two-cycles in K3 will be of order $\sim \alpha'f(Q_1/Q_5)$ for
certain functions $f$. Altogether this shows that we can neglect 
all branes wrapped on various cycles in K3. The KK fields from
$S^3$ will have masses of order $m^2 \alpha'\sim 1/\ell^2$, whereas
massive string states will have masses of order $m^2 \alpha' \sim
1$. KK states from K3 will have masses of order $m^2 \alpha' \sim
\sqrt{Q_1/Q_5}$ and of order $m^2 \alpha' \sim f(Q_1/Q_5)$.

S-duality of type IIB string theory sends $\exp(\phi)\rightarrow
\exp(-\phi)$, and $g_{\mu\nu}\rightarrow \exp(-\phi) g_{\mu\nu}$. 
After a trivial change of the $U$-coordinate the metric becomes
\be
\frac{ds^2}{\alpha'}  =  \frac{U^2}{Q_5}
(-dt^2 + (dx^5)^2) + \frac{Q_5}{U^2}dU^2
+ Q_5 d\Omega_3^2 
 + \frac{1}{g_6 \sqrt{v}} ds^2_{K3}
\ee
This metric corresponds to an exact 
string background, which consists of a K3 piece, a level $Q_5$ $SU(2)$
WZW theory and a level $Q_5$ $Sl(2,{\bf R})$ piece.
The dilaton
$\exp(2 \phi)=Q_5/g_6^2 Q_1$ blows up in the limit we are interested in,
so we expect that only certain BPS-protected quantities can
be meaningfully compared. In particular, one should be able
to recover the chiral primaries of the conformal field theory
from this string background. A recent result in this direction is \cite{wzw2}
where the stringy exclusion principle\footnote{The stringy exclusion
principle can also be understood classically in terms of large
gauge transformations \cite{marttalk,groo}.}
 of \cite{mastr}
is related to a unitary truncation of the $Sl(2,{\bf R})$ WZW spectrum.
It would be interesting to explore the relation between the
conformal field theory and this exact string background in 
more detail.

\section{Six-Dimensional Supergravity}

\subsection{Symmetries and the anti-de Sitter supergroup}

As is clear from the discussion in the previous section, the
low-energy excitations of the conformal field theory of the
D1-D5 brane system should be compared to the 
KK spectrum of six-dimensional
supergravity compactified on ${\rm AdS}^3 \times S^3$. 
Although initially our focus will be on the chiral six-dimensional
theory obtained by putting type IIB supergravity on K3,
later we want consider various six-dimensional
theories, with various numbers of matter multiplets. The supersymmetry
generators in six dimensions are Weyl spinors of a fixed chirality, 
and the total number of chiral (antichiral) spinors will be denoted by
$2 n_L$ ($2 n_R$). Six-dimensional theories exist for various values of 
$n_L$ and $n_R$. The minimal theory has $n_L=1,n_R=0$, and appears
for instance in heterotic and type I compactifications on $K3$. 
Type IIA on $K3$
yields a 6d theory with $(n_L,n_R)=(1,1)$, type IIB on $K3$ yields
a 6d theory with $(n_L,n_R)=(2,0)$, and type IIA or IIB on $T^4$ yields
a theory with $(n_L,n_R)=(2,2)$. Although other values of $(n_L,n_R)$
are possible as far as the existence of a supersymmetry algebra is
concerned, there is only one other algebra whose representations include
a graviton, namely $(n_L,n_R)=(2,1)$. The corresponding supergravity
theory is presumably anomalous \cite{afk}, but we can still consider
its KK spectrum, as that requires only classical supergravity. 
A convenient table of the various superalgebras and their multiplets
is given in \cite{stra}.

Six-dimensional supergravity has an $Sp(n_L) \times Sp(n_R)$ 
global symmetry. These symmetries do not play an important
role in the rest of this paper, and will therefore be ignored.
In any case, it is straightforward to keep track of the 
global symmetries at each step and determine how various quantities
transform under them. Besides these global symmetries, 6d supergravity
on $S^3 \times AdS_3$ has an $SO(4)\times SO(2,2) \simeq
SU(2) \times SU(2) \times Sl(2,{\bf R}) \times Sl(2,{\bf R})$
group of isometries. These are part of an anti-de Sitter supergroup
$G=G_L \times G_R$, where both $G_L$ and $G_R$ contain 
$SU(2)\times Sl(2,{\bf R})$. $G_L$ and $G_R$ are global symmetries
of the left and right-moving degrees of freedom of the two-dimensional
conformal field theory. The simple supergroups that contain $Sl(2,{\bf R})$
were classified in \cite{gst}. The only simple supergroups whose
bosonic part is $SU(2)\times Sl(2,{\bf R})$ are $Osp(3|2,{\bf R})$
and $SU(1,1|2)$, but the first one (that actually contains 
an $O(3)$ subgroup) can be easily ruled out by looking at the
transformation properties of the supercharges. All supercharges
transform in the spinor representation of $SO(4)$ and 
should therefore be in the spin-1/2 representation of $SU(2)$.
However, the fermionic generators of $Osp(3|2,{\bf R})$
are in the vector representation of $O(3)$. Thus the only
possible supergroup is $SU(1,1|2)$. 

The $S^3 \times AdS_3$ background is maximally supersymmetric. 
For $n_L+n_R>1$, the supercharges transform in the
$2 (n_L+n_R)(({\bf 2},{\bf 1})\oplus ({\bf 1},{\bf 2}))$ representation
of $SO(4)$, and for $n_L=1$, $n_R=0$ as $4({\bf 2},{\bf 1})$. 
In general, only a subset of the supercharges will close into
the $SO(4) \times SO(2,2)$ generators, and these will
form the AdS supergroup $G_L \times G_R = SU(1,1|2) \times
SU(1,1|2)$ for $n_L+n_R>1$ and $G_L \times G_R = SU(1,1|2)
\times Sl(2,{\bf R}) \times SU(2)$ for $n_L=1,n_R=0$.

It is very useful to know the anti-de Sitter supergroups.
It helps organize the KK spectrum of supergravity,
which should fall in representations of the supergroup, and
it also tells us something about the chiral algebra of the
conformal field theory living at the boundary. 

\subsection{The chiral algebra at the boundary}

The relation between the chiral algebra at the boundary and
$G_L$, $G_R$ arises as follows. At low energies the supergravity theory on
$AdS_3$ is described by the difference between two Chern-Simons
theories \cite{towncs,towncs2,witcs},
\be \label{p1}
S=\frac{k}{4\pi} \int (A \wedge dA + \frac{2}{3} A\wedge A \wedge A) -
\frac{k}{4\pi} \int (\bar{A} \wedge d\bar{A} + \frac{2}{3} 
\bar{A}\wedge \bar{A} \wedge \bar{A})
\ee
where $A$ is a connection for $G_L$ and $\bar{A}$ a connection for
$G_R$. Here, $k=Q_1 Q_5$ for the D1-D5 brane system. More generally,
$k=\ell^4/G_6$, where $G_6$ is a dimensionless six-dimensional
Newton's constant that appears in the six-dimensional action as
$S=\frac{1}{8\pi^3 \alpha'^2 G_6} \int d^6 x \sqrt{-g} R$. For
the D1-D5 brane system, $G_6=g^2/v=g_6^2$. 

To see the relation between this Chern-Simons theory and the chiral
algebra on the boundary, we generalize the arguments of 
\cite{bancs,bancs2,hencs,mart,strom}, and in particular of
\cite{bancs3,hencs2}. Since the connections $A,\bar{A}$ contain
the three-dimensional spin-connection and vielbein, we have to impose
a suitable boundary condition on them in order to describe $AdS_3$.
The vielbein and spin-connection live in the $Sl(2,{\bf R})$ subgroup
of $G_L$. Denote by $Sl(2,{\bf R}) \stackrel{j}{\hookrightarrow} 
G_L$ the corresponding embedding. Then the boundary conditions
are that to leading order in $r=U/\ell^2$
\be
A_i \sim b^{-1} \partial_i b + b^{-1} A_i^{(0)} b
\ee
where 
\be
b=j\left( \begin{array}{cc} \sqrt{r} & 0 \\ 0 & 1/\sqrt{r} \end{array}
\right)
\ee
and $A^{(0)}_r=A^{(0)}_-=0$, $x^{\pm}=t\pm x^5$, and
\be
A^{(0)}_+ = j(T^-) + T(x^+,x^-).
\ee
Here, $T^+,T^0,T^-$ is a basis for $sl(2,{\bf R})$ and $T(x^+,x^-)$
is an arbitrary 
$x^{+},x^{-}$-dependent
element of the Lie algebra of $G_L$ subject to the
condition $[j(T^+),T(x^+,x^-)]=0$. 
The component of $T(x^+,x^-)$ proportional to $j(T^+)$ is
the stress energy tensor.
For instance, for $Sl(2,{\bf R})$
$A^{(0)}_+=\left( \begin{array}{cc} 0 & T(x^+,x^-) \\ 1 & 0 \end{array}
\right)$. The constraints imposed on $A^{(0)}_+$ are exactly
the same constraints one imposes on the currents of WZW theory
when one performs Hamiltonian reduction based on embeddings of 
$sl_2$ \cite{alsh,balog,beroog,btvd}. The relation between constrained
currents and 2d gravity \cite{p,kpz} and Chern-Simons theory \cite{hv,sc}
is well-known. The generators of the chiral algebra 
of the boundary theory are in one-to-one correspondence with
the components of $T(x^+,x^-)$. The symmetries they generate are
precisely the gauge transformations that preserve the form
of $A^{(0)}_+$, and from the form of the gauge transformations 
one can immediately read off their Poisson brackets. In particular,
the central charge is equal to $c=6k$. 

Thus, to summarize, the left-moving 
chiral algebra of the conformal field theory
at the boundary is given by the Hamiltonian reduction of the current
algebra based on $G_L$, and similarly for the right-movers. Conversely,
$G_L$ is obtained by keeping only the modes $A_m$ with $|m|<\Delta$
for each spin-$\Delta$ generator of the chiral algebra. 
Below we list the supergroups listed in \cite{gst} and the chiral
algebras one obtains from them by Hamiltonian reduction.

\begin{table}[h]
\caption{2d anti-de Sitter supergroups and their chiral algebras.
See \cite{alg1,alg2,alg3,alg4,alg5,alg6} and references therein 
for more details.  }
\begin{center}
\begin{tabular}{|l|l|}
\hline
$Osp(N|2,{\bf R})$ & $O(N)$ extended superconformal algebras;\\
 &  $N=1$ and $N=2$ are the usual $N=1,2$ superconformal algebras \\
$SU(N|1,1)$ & $U(N)$ extended superconformal algebras \\
$SU(2|1,1)$ & ``small'' $N=4$ algebra \\
$Osp(4^{\ast}|2N)$ & $Sp(N)$ extended superconformal algebras \\
$G(3)$ & octionic $N=7$ algebra \\
$F(4)$ & octionic $N=8$ algebra \\
$D(2,1,\alpha)$ & ``large'' $N=4$ algebra ${\cal A}_{\gamma}$ \\
\hline
\end{tabular}
\end{center}
\end{table}

As we already pointed out, the algebra of interest for our case
is $SU(2,1|1)$, in agreement with the fact that the boundary CFT
is a $N=(4,4)$ of $N=(4,0)$ conformal field theory.

\subsection{$AdS$ masses and conformal weights}

The relation between the mass in $AdS$ and the conformal weights at
boundary was already discussed for scalars in section~4.1 of \cite{mastr}.
Here we extend this discussion to include fields of arbitrary
spin $s$ on $AdS_3$. The first step is to map the plane at $U=\infty$
to a cylinder, after which the $AdS_3$ part is described by the 
metric 
\be \frac{ds^2}{\ell^2} = -\cosh^2 \rho \, d\tau^2 
+ \sinh^2 \rho d \phi^2 + d \rho^2.
\ee
Including the ``isospin part'', the Virasoro generators are 
($u=\tau + \phi$, $v=\tau-\phi$)
\bea \label{expl1}
L_0 & = & i \partial_u \nonumber \\
L_{-1} & = & i e^{-i u} 
\left( \coth 2\rho \,\partial_u -
\frac{1}{\sinh 2\rho} \partial_v + \frac{i}{2} \partial_{\rho}
-\frac{i}{2} s\,  \coth\rho \right) \nonumber \\
L_1 & = & i e^{i u} 
\left( \coth 2\rho \,\partial_u -
\frac{1}{\sinh 2\rho} \partial_v - \frac{i}{2} \partial_{\rho}
+\frac{i}{2} s\,  \coth\rho \right) 
\eea
and similarly for $\bar{L}_{0,\pm 1}$ with $u\leftrightarrow v$ and
$s \rightarrow -s$. Primary fields satisfying $L_1\,\psi= \bar{L}_1
\, \psi =0$, $L_0 \,\psi = h \psi$ and $\bar{L}_0 \psi = \bar{h}
\psi$ exist only if $s=h-\bar{h}$ and
are then given by
\be
\psi \sim \frac{e^{-ihu-i\bar{h} v} }{(\cosh \rho)^{h+\bar{h}}}
\ee
By evaluating the sum of the Casimirs of the two $Sl(2,{\bf R})$'s
using the explicit expressions (\ref{expl1}) we find that
for the primary field
$\psi$
\be 
(2h(h-1) + 2\bar{h} (\bar{h}-1) )\psi = \ell^2 \Box \psi + s^2 \coth^2 \rho\,
\psi
\ee
where $\Box$ is the Laplacian on scalar fields. This can be
rewritten as
\be \label{lapeq}
(\Box + \frac{s^2}{\ell^2 \sinh^2 \rho})\psi = m^2 \psi
\ee
with
\be
\label{massrel} 
\ell^2 m^2 = 2h(h-1) + 2\bar{h} (\bar{h}-1) - s^2 = (h+\bar{h})
(h+\bar{h}-2) .
\ee
For large $\rho$, the angular momentum part in (\ref{lapeq}) decouples,
and $m$ is what we would like to call the $AdS$ mass. 
Equation (\ref{massrel}), together with $s=h-\bar{h}$, completely determine
the spin and mass on $AdS_3$ in terms of the conformal weights
at the boundary, and vice versa.

\section{The Kaluza-Klein Spectrum}

\subsection{The spectrum of harmonics}

To determine the KK spectrum of some 6d supergravity compactified on
$S^3 \times AdS^3$, we need to expand every field in harmonics on $S^3$,
insert this into the linearized field equations, and diagonalize the remaining
equations on $AdS^3$. From these field equations one can read off the
masses of the various KK excitations on $AdS^3$.
This is a rather complicated procedure, which was 
done for various ten and eleven-dimensional
cases in \cite{kk1,kk2,kk3,kk4,kk5,kk6}, and recently for 6d $(2,0)$
supergravity in \cite{sez1}. To determine precisely the $AdS$ mass
corresponding to each harmonic requires a 
rather precise knowledge of
the field equations, but to determine which harmonics appear
requires only a knowledge of the field content of the theory. 

In fact, it is rather easy to find the complete spectrum of 
harmonics in the theory. The sphere $S^3$ is the homogeneous space
$SO(4)/SO(3)$, and the harmonics that appear on homogeneous spaces
were discussed in \cite{ss}. Any field on the theory can be decomposed
as the sum of products of fields living on $AdS^3$ and $S^3$. Each field on
$S^3$ transforms in some
representation $R_{4}$ of the isometry group $SO(4)$, but in addition transforms
in some representation $R_3$ of the local Lorentz group $SO(3)$ of $S^3$. 
According to \cite{ss}, the only representations $R_{4}$ of $SO(4)$
that appear in the harmonic expansion are those that contain $R_3$ in the
decomposition of $R_{4}$ in $SO(3)$ representations. 
For example, a vector field on $S^3$ transforms in the ${\bf 3}$ of
$SO(3)$. This $SO(3)=SU(2)$ is the diagonal $SU(2)$ subgroup of
$SO(4)=SU(2) \times SU(2)$. If we decompose a representation
$({\bf 2j_1+1},{\bf 2j_2+1})$ of $SO(4)$ in terms of $SO(3)$ representations 
and require that the result contains ${\bf 3}$, we find that 
$|j_1-j_2|$ has to be zero or one. Thus, a vector field on $S^3$
gives rise to
the harmonics that transform as $({\bf m},{\bf m-2})$,
 $({\bf m},{\bf m})$ and $({\bf m},{\bf m+2})$
under $SO(4)$.

The various $SO(3)$ representations that appear when decomposing a field in
pieces living on $AdS^3$ and $S^3$ can be found as follows. Since ultimately
we are interested in the solutions of the field equations modulo gauge 
invariance, the degrees of freedom contained in some six-dimensional
field are labeled by a representation of the little group $SO(4)$. 
The fields can depend arbitrarily on the coordinates of $S^3$, and
the local Lorentz
group $SO(3)$ of the three-sphere should therefore be
identified with an $SO(3)$ subgroup
of the little group $SO(4)$. Thus, all $SO(3)$ representations are found
by decomposing the representation of the little group in representations of 
$SO(3)$\footnote{More generally, consider some $p+q$-dimensional theory
compactified on $S^p \times AdS_q$. Any field $\phi$ transforming in some
representation $R$ of the little
group $SO(p+q-2)$ can be decomposed as $\oplus_i n_i S_i \otimes T_i$,
where $S_i$ and $T_i$ are irreducible representations of $SO(p)$
and $SO(q-2)$. If $U_{i,r}$, $r=1,\ldots$ is the set representations
of $SO(p+1)$ that contain $S_i$ when decomposing it into $SO(p)$ representations,
then the set of harmonics for $\phi$ is $\oplus_{i,r} n_i C_{i,r}$}.

As an example, consider the metric in six dimensions. The graviton 
transforms in the $({\bf 3},{\bf 3})$ of the little group $SO(4)$,
so that the relevant $SO(3)$ representations are ${\bf 1} + {\bf 3} + 
{\bf 5}$. For each of these we find the corresponding harmonics as above,
and altogether we find that the total set of harmonics coming from the
graviton transforms as
\be
\oplus_m \left( ({\bf m} ,{\bf m-4}) + 2({\bf m},{\bf m-2})
 + 3({\bf m},{\bf m}) + 2({\bf m},{\bf m+2}) + ({\bf m},{\bf m+4}) \right)
\ee

As another example, consider $(n_L,n_R)=(2,0)$ supergravity with $n_T$
tensor multiplets. The theory is anomalous unless $n_T=21$ which is
the theory obtained from type IIB on K3, but for the purpose of counting
harmonics we can take $n_T$ arbitrary. The gravity multiplet consists
of a graviton, five self-dual two-forms and four gravitinos, and
the tensor multiplet of one anti self-dual two-form, four fermions and
five scalars. The total representation of the little group is \cite{stra}
\be
({\bf 3},{\bf 3}) + 
4 ({\bf 2},{\bf 3}) + 
5 ({\bf 1},{\bf 3}) + 
n_T ({\bf 3},{\bf 1}) + 
4n_T ({\bf 2},{\bf 1}) + 
5 n_T({\bf 1},{\bf 1})
\ee
which yields for the total set of harmonics
\bea \label{harm11}
& & \oplus_m \left( ({\bf m },{\bf m\pm 4 }) + 4({\bf m },{\bf m\pm 3 }) +
 (n_T+7)({\bf m},{\bf m \pm 2})
\right. \nonumber \\ & & \qquad \left. +
 4(n_T+2) ({\bf m },{\bf m \pm 1 }) + (6 n_T + 8)({\bf m },{\bf m }) \right) .
\eea

\subsection{Representations of the AdS supergroup}

Having determined the total set of harmonics, we would like to extract
from them the spectrum of the fields of the SCFT living at the
boundary of $AdS^3$. In principle we could first determine the masses and
spins of the various KK modes on $AdS^3$ from the field
equations, and then relate these masses to
the scaling dimensions using (\ref{massrel}).
However, it is also possible to avoid this
and to obtain the scaling dimensions and $U(1)$ weights at the boundary
using representation theory, if there are sufficiently many
supersymmetries.  The point is that the complete KK spectrum
should fall into representations of the anti-de Sitter supergroup $G_L \times
G_R$, ,
whose generators include the generators $L_0$ and $J_0$ of the
boundary (super)conformal algebra. If our knowledge of the harmonics
is enough to completely determine the set of representations 
of the anti-de Sitter supergroup that
appear, we can simply read of the various values of $L_0$ and $J_0$
that appear in the boundary CFT, and in particular we can determine
the set of chiral primaries. In \cite{kk2} this strategy was used
to determine the KK spectrum of IIB supergravity on $S^5 \times AdS^5$.
In that case, the only representations of the relevant anti-de Sitter
supergroup $U(2,2|4)$ that appear are in so-called short representations.
This follows from the fact that short representations are the only
ones having spins of at most two as required for a theory of supergravity.
Although this argument is restricted to theories with 32 supersymmetries,
we will find that also for the six-dimensional theories with 16 and 8
supersymmetries the only multiplets appearing in the KK reduction are
short multiplets. 

In the case of $AdS^3$, the relevant anti-de Sitter supergroup is not
simple but the product of two groups $G_L \times G_R$. 
The possible simple supergroups for $G_{L,R}$ are reproduced in
Table~1. 
For a discussion of these algebras and their representations, see
\cite{gst}. The representations are constructed in \cite{gst}
using the oscillator method \cite{osc1}, which means that one expresses the generators
of the algebra in terms of a set of free bosonic and harmonic oscillators and
then uses the Fock space of these oscillators to construct representations
of the algebra. Short representations are obtained by acting with
the generators on the Fock vacuum, and are
labeled by the number of oscillators in terms of which the generators
are constructed. 

In our case, we are mainly interested in the representations of
$SU(2|1,1)$. Although the structure of short representations of
this superalgebra is well-known (it parallels the representation
theory of the $N=4$ superconformal algebra), we will rederive it
here using the oscillator method because that method generalizes
to other algebras and other dimensions.

For $SU(2|1,1)$, we introduce $\{\vec{a}_i,\vec{a}^{\dagger}_i,
\vec{\psi}_i,\vec{\psi}_i^{\dagger} \}$, where $i=1,2$, $\vec{a}_i$ 
are $k$-component vectors of
bosonic creation and annihilation operators and
$\vec{\psi}_i$ are $k$-component vectors of fermionic
creation and annihilation operators.
The brackets are
\be
[a_i(p),a_j(q)^{\dagger}]=\delta_{p,q}\delta_{i,j},\qquad
\{ \psi(p)_i, \psi(q)_j^{\dagger} \} = \delta_{p,q} \delta_{i,j}
\ee

The lowering operators of $SU(2|1,1)$ are
\be 
L_- = \{ \vec{a}_1 \ccdot \vec{a}_2 ,\vec{a}_1 \ccdot \vec{\psi}_1 \pm
\vec{a}_2 \ccdot \vec{\psi}_2, \vec{\psi}_1 \ccdot  \vec{\psi}_2 \} ,
\ee
the generators of the maximal compact subsuperalgebra are
\be
L_0 = \{ \vec{a}_1 \ccdot\vec{a}^{\dagger}_1  +
\vec{a}^{\dagger}_2 \ccdot \vec{a}_2  ,
          \vec{\psi}^{\dagger}_1 \ccdot \vec{\psi}_1 -
  \vec{\psi}_2 \ccdot  \vec{\psi}^{\dagger}_2,
           \vec{a}_1 \ccdot\vec{\psi}^{\dagger}_2 \pm
  \vec{a}_2\ccdot \vec{\psi}^{\dagger}_1,
           \vec{a}^{\dagger}_1 \cdot\vec{\psi}_2 \pm
  \vec{a}^{\dagger}_2\ccdot \vec{\psi}_1\} ,
\ee
and the raising operators are
\be
L_+ = \{ \vec{a}^{\dagger}_1 \ccdot \vec{a}^{\dagger}_2 ,
\vec{a}^{\dagger}_1 \ccdot \vec{\psi}^{\dagger}_1 \pm
\vec{a}^{\dagger}_2 \ccdot \vec{\psi}^{\dagger}_2, 
\vec{\psi}^{\dagger}_1 \ccdot  \vec{\psi}^{\dagger}_2 \} .
\ee
A single irrep of  $SU(2|1,1)$ contains several
different irreps of $Sl(2,{\bf R}) \times SU(2)$. The fermionic
operators $Q^{\pm}=\vec{a}^{\dagger}_1 \cdot \vec{\psi}^{\dagger}_1 \pm
\vec{a}^{\dagger}_2 \cdot \vec{\psi}^{\dagger}_2$ map
between the various irreps of  $Sl(2,{\bf R}) \times SU(2)$.
The algebra $Sl(2,{\bf R})$ consists of the Virasoro generators
$L_{-1},L_0,L_{+1}$. These generators are given by
\bea
L_{-1} & = &   \vec{a}^{\dagger}_1 \ccdot \vec{a}^{\dagger}_2 \\{}
L_{0} & = & \frac{1}{2}( \vec{a}_1 \ccdot\vec{a}^{\dagger}_1  +
\vec{a}^{\dagger}_2 \ccdot \vec{a}_2 )
 \\{}
L_{+1} & = & \vec{a}_1 \ccdot \vec{a}_2  .
\eea

It will also be convenient to identify the $SU(2)$ 
subgroup 
\bea 
J^- & = &  \vec{\psi}_2 \ccdot  \vec{\psi}_1 \\{}
J^0 & = & \vec{\psi}^{\dagger}_1 \ccdot \vec{\psi}_1 -
  \vec{\psi}_2 \ccdot  \vec{\psi}^{\dagger}_2 \\{}
J^+ & = & \vec{\psi}^{\dagger}_1 \ccdot  \vec{\psi}^{\dagger}_2 .
\eea
Here, $J^0$ is normalized so that it agrees with the usual
definition of $J^0$ in an $N=2$ superconformal algebra.

As in \cite{gst}, we find that the short multiplets consist of
\be \label{short}
\begin{array}{rccc}
{\rm states} & j & j' &  L_0 \\
|0\rangle & k/2 & 0 & k/2 \\
Q^{\pm} |0\rangle & (k-1)/2 & 1/2 &  (k+1)/2 \\
Q^{\pm} Q^{\pm} |0\rangle & (k-2)/2 & 0 & (k+2)/2 
\end{array}
\ee
where $j$ is the spin of $SU(2)$, $j'$ is the spin with respect
to the global $SU(2)$ automorphism group under which $Q^{\pm}$ is a
doublet, and $L_0$ is the conformal weight of the ground
state. Thus a short multiplet of $SU(2|1,1)$
consists of four representations of $Sl(2,{\bf R}) \times SU(2)$.
Not surprisingly, the structure is exactly the same as that
of a representation of the $N=4$ superconformal algebra whose
hightest weight state is a chiral primary. The representation
(\ref{short}) contains indeed precisely one chiral primary,
namely the vacuum $|0\rangle$ which has $J^0$ eigenvalue $k$ and
$L_0$ eigenvalue $k/2$.  It is easy to see that only
short multiplets contain a chiral primary field, as one would expect.

An example of a long multiplet is the one built out of the ground
states $\psi_1^{\dagger}(1) |0\rangle$, $a_2^{\dagger}(1) |0\rangle$.
The content of this multiplet reads
\be \label{long}
\begin{array}{rccc}
{\rm states} & j & j' & L_0 \\
a_2^{\dagger}(1) |0\rangle & k/2 & 0 & (k+1)/2 \\
\psi_1^{\dagger}(1) |0\rangle & (k-1)/2 & 0 & k/2 \\
Q^{\pm} a_2^{\dagger}(1)|0\rangle & (k-1)/2 & 1/2 & (k+2)/2 \\
Q^{\pm}\psi_1^{\dagger}(1)|0\rangle & (k-2)/2 & 1/2 & (k+1)/2 \\
Q^{\pm} Q^{\pm} a_2^{\dagger}(1)|0\rangle & (k-2)/2 & 0 & (k+3)/2 \\
Q^{\pm} Q^{\pm} \psi_1^{\dagger}(1) |0\rangle & (k-3)/2 & 0 & (k+2)/2 
\end{array}
\ee

Let us now come back to the compactifications of 6d supergravities on
$S^3 \times AdS^3$ with $n_L+n_R>1$. 
The complete KK spectrum should organize itself as
representations of 
$G_L \times G_R=SU(2|1,1) \times SU(2|1,1)$.
We will assume that all KK states will fall into short representations
of $SU(2|1,1) \times SU(2|1,1)$. If we denote the short
representation given in (\ref{short}) by ${\bf k+1}$, then any short
representation of the product group is of the form $({\bf k+1},{\bf k'+1})_S$.
The subscript $S$ has been included in order to avoid confusion with
representations $({\bf m},{\bf m'})$ of $SO(4)$, the group 
of rotations of the three-sphere. The idea is now to take a set of
short multiplets, decompose these into $SO(4)$ representations, and
compare the result to the set of $SO(4)$ representations obtained
from the KK analysis of supergravity described in the previous section.
Requiring that the two agree will give us the set of short multiplets,
and from (\ref{short}) we then obtain the spectrum of highest weight
states of the CFT, and in particular the set of chiral primaries of
the CFT. Supergravity yields only ``single particle'' states of the
conformal field theory at the boundary. The full spectrum is obtained
by taking arbitrary products of these single particle states. 

It turns out that in all cases with $n_L+n_R>1$ 
the spectrum of short multiplets is of the form
\be \label{e1}
\oplus_m \left( t_{2} ({\bf m},{\bf m\pm 2})_S +
t_{1} ({\bf m},{\bf m\pm 1})_S +
t_{0} ({\bf m},{\bf m})_S  \right)
\ee   
The reason that the difference between the two integers is at most two
is because supergravity has fields of spin at most two. A larger
difference would correspond to higher spin fields. 
Using the structure of the short multiplet (\ref{short}), we can decompose
each $({\bf m},{\bf m'})_S$ in representations of $SO(4)$,
\be
({\bf m},{\bf m'})_S =\oplus_{i=0}^2 \oplus_{j=0}^2 
\left( \begin{array}{c} 2 \\ i \end{array} \right)
\left( \begin{array}{c} 2 \\ j \end{array} \right)
({\bf m-i},{\bf m'-j})
\ee
which shows that (\ref{e1}) is equivalent to the following spectrum of $SO(4)$
representations
\bea \label{decomp}
& & \oplus_m \left(
t_2 ({\bf m},{\bf m\pm 4}) + (4t_2 + t_1) ({\bf m},{\bf m\pm 3})
+ (6t_2 + 4t_1 + t_0) ({\bf m},{\bf m \pm 2}) \right. \nonumber \\{} & & \qquad \left.
+ (4t_2 + 7t_1 + 4t_0) ({\bf m},{\bf m \pm 1})
+ (2t_2 + 8t_1 + 6t_0) ({\bf m},{\bf m}) \right)
\eea

Before proceeding, we will first discuss the spectrum of the conformal
field theory in some more detail, and then compare (\ref{decomp})
to the KK spectrum of various supergravities.

\section{The $K3^N/S_N$ Conformal Field Theory}

The conformal field theory of the D1-D5 system with the D5 branes
wrapped on $K3$ has been conjectured to be described by a deformation
of the supersymmetric sigma model whose target space is the the 
orbifold $K3^N/S_N$ \cite{va1,vast}. In order to compare the spectrum
obtained from the KK reduction to that of the conformal field theory,
we will need to know in some detail the spectrum of this conformal
field theory. 

The most robust set of states in the conformal field theory are
the states which are chiral primary both for the left and right movers.
Since these states are in ultrashort multiplets and their conformal
weights satisfy a BPS bound, their spectrum is
independent of any perturbation of the conformal field theory,
and can be conveniently encoded in terms of the generalized 
Poincar\'e polynomial
\be
P_{t,\bar{t}}  = \tr( t^{J_0} \bar{t}^{\bar{J}_0} )
\ee
where the trace is taken over the space of chiral primaries only.
In case the superconformal field theory is a supersymmetric sigma
model with target space $M$, the Poincar\'e polynomial equals
\be
P_{t,\bar{t}} = \sum_{p,q} h_{p,q} t^p \bar{t}^q
\ee
where $h^{p,q}$ are the Betti numbers of $M$ \cite{witten1}. 
The Poincar\'e polynomial of a resolution of $K3^N/S_N$ called the
Hilbert scheme of $N$ points on $K3$
was computed in \cite{goso} and has generating function
\be \label{Poincare}
\sum_{N\geq0} Q^N
P_{t,\bar{t}}(K3^N/S_N) = 
\prod_{m=1}^{\infty} \prod_{p,q}
\left( 1 + (-1)^{p+q+1} Q^m t^{p+m-1}
\bar{t}^{q+m-1} \right)^{ (-1)^{p+q+1} h^{p,q} }.
\ee

An alternative derivation of this result uses standard orbifold 
conformal field theory. The Hilbert space of $S_N$ orbifolds
can be decomposed in terms of Hilbert spaces of ${\bf Z}_n$
orbifolds as in \cite{dmvv}. According to 
the discussion of ${\bf Z}_n$ orbifolds in
\cite{orb}, a state with conformal weight $h$ and $U(1)$
weight $q$ in the original CFT $M$ gives rise to various states in
the orbifold CFT $M^n/{\bf Z}_n$. In the untwisted sector we get
a state with $h'=nh$ and $q'=nq$, and in the twisted sector states
with $q'=q$ and $h'=\frac{h+m}{n}+\frac{c}{24}\frac{n^2-1}{n}$,
where $m$ is some nonnegative integer. 

Consider now a supersymmetric sigma model with target space a complex
K\"ahler manifold $M$ of complex dimension $d$, and central charge
$c=3 d$. A chiral primary with $h=p/2$ and $q=p$ in the left-moving NS
sector corresponds via spectral flow to a Ramond ground state
with $h=d/8$ and $q=p-d/2$. According to the discussion above,
the untwisted sector of $M^n/{\bf Z}_n$ contains a Ramond ground
state with $h=nd/8$ and $q=n(p-d/2)$, whereas the twisted sector contains
a Ramond ground state with $h=nd/8$ and $q=p-d/2$ (the case $m=0$).
This corresponds to chiral primaries with $(h,q)=(np/2,np)$ and
$(h,q)=((p+\frac{d}{2}(n-1))/2,p+\frac{d}{2}(n-1))$ in the NS sector.
The complete Poincare polynomial is now determined by combining these
latter states with their right moving counterparts, and by subsequently
writing down a second quantized partition function for these generators
\cite{dmvv}. This leads to
\be \label{Poincare2}
\sum_{N\geq 0} Q^N P_{t,\bar{t}}(M^N/S_N) = \prod_{m=1}^{\infty} \prod_{p,q}
\left( 1 + (-1)^{p+q+1} Q^m t^{p+\frac{d}{2}(m-1)}
\bar{t}^{q+\frac{d}{2}(m-1)} \right)^{ (-1)^{p+q+1} h^{p,q} } .
\ee
For $d=2$ we indeed recover (\ref{Poincare}).

Interestingly, $ P_{t,\bar{t}}(M^N/S_N)$
 does have a well-defined $N\rightarrow \infty$
limit. Because it has a single factor of $(1-Q)^{-1}$, (\ref{Poincare2})
is of the form
\be
a_0  + (a_0 + a_1)Q + (a_0+a_1+a_2) Q^2 + \ldots = 
(1-Q)^{-1} (a_0 + a_1 Q + a_2 Q^2 + \ldots )
\ee
Thus the $N\rightarrow \infty$ limit is obtained by extracting the factor
of $(1-Q)^{-1}$ and taking  $Q\rightarrow 1$,
\be \label{e56}
P_{t,\bar{t}}(M^{\infty}/S_{\infty}) = 
\lim_{Q\rightarrow 1} (1-Q) \sum_{N\geq 0} Q^N P_{t,\bar{t}}(M^N/S_N).
\ee
For $d=2$, this is the partition function for a set of unconstrained
bosonic and fermionic oscillators. If we define $n_{\Delta}=\sum_{p}
h^{p+\Delta,p}$, then for sufficiently large $m$ there will be
$n_{\Delta}$ oscillators of degree $(m+\Delta,m)$. The oscillators
are bosonic (fermionic) depending on whether $\Delta$ is even (odd).
The only exception is that there are only $h^{1,0}$ generators of
degree $(1,0)$, $h^{0,1}$ of degree $(0,1)$, and $h^{0,0}+h^{1,1}$ of
degree $(1,1)$. In particular, for K3 there are only bosonic generators,
one of degree $(m,m+2)$ and one of degree $(m+2,m)$ for $m\geq0$,
22 of degree $(m,m)$ for $m>1$ and 21 of degree $(1,1)$.

The spectrum of left and right-moving chiral primaries is not
the only part of the spectrum which is independent of marginal
deformations of the theory. A more general object with this
property is the elliptic genus, which can only change if a phase
transition occurs. The elliptic genus is defined by
\be
Z(\tau,z)=\tr_{RR} (-1)^F q^{L_0-c/24} \bar{q}^{\bar{L}_0-c/24} y^{J_0}
\ee
with $q=e^{2\pi i\tau}$ and $y=e^{2\pi i z}$, and the trace is over
the Ramond sector of the Hilbert space \cite{ell1,ell2,ell3}.
The elliptic genus for $K3$ was considered in \cite{ell4}, and
its explicit form is \cite{ell5}
\be
\label{ellgenus}
Z(\tau,z) \equiv\sum_{m,l} c(m,l) q^m y^l= 24 \left(
\frac{\theta_3(\tau,z)}{\theta_3(\tau,0)}\right)^2 - 
2 \frac{\theta_4(\tau,0)^4 - \theta_2(\tau,0)^4}{\eta(\tau)^4}
\left(\frac{\theta_1(\tau,z)}{\eta(\tau)} \right)^2.
\ee
With this definition of $c(m,l)$, the elliptic genus of $K3^N/S_N$ 
has generating function \cite{dmvv}
\be
\sum_{N\geq 0}  p^N Z(K3^N/S_N;\tau,z) = 
\prod_{n>0,m\geq0,l} \frac{1}{(1-p^n q^m y^l)^{c(nm,l)}}.
\ee
The first few terms in the elliptic genus of $K3^N/S_N$ for $N>6$
read\footnote{Notice that the elliptic genus diverges
for $N\rightarrow \infty$, so
that we cannot use it to count states in the strict supergravity limit.}
\bea \label{ellexp}
Z(K3^N/S_N;\tau,z) & = & ((N+1)y^{-N} + \ldots) \nonumber \\
                              & & +q ((22N-2)y^{-N-1}
 + (464N-592) y^{-N} + \ldots) \nonumber \\
                              & & + q^2 ((277N-323) y^{-N-2} 
+ (5652N-13716) y^{-N-1} \nonumber  \\ & & \qquad
+ (67131N-244053) y^{-N}+\ldots ) \nonumber \\
                             & & + {\cal O}(q^3)
\eea
Translating (\ref{ellexp}) back to the NS-NS sector, we find for example that
the orbifold CFT has $67131N-244053$ states of the form
$|q,h\rangle_L \otimes |q',h'\rangle_R=
|0,2\rangle_L \otimes |q',q'/2\rangle_R$ where
we should count the states weighted with the sign $(-1)^{q'}$. 
To count how many states of this form are descendants of chiral
primaries, we need to know the first few terms in 
the Poincar\'e polynomial $P_{t,-1}(K3^N/S_N)$. The coefficient
in front of $t^q$ counts the number of chiral primaries of the form
$|q,q/2\rangle_L \otimes |q',q'/2\rangle_R$, weighted with $(-1)^{q'}$. For 
sufficiently large $N$ we find
\bea \label{coho}
P_{t,-1}(K3^N/S_N) & = & (N+1) -(22N-2) t + (277N-323)t^2
-(2576N-5752)t^3 \nonumber \\
& & + (19574N-64474)t^4 + {\cal O}(t^5).
\eea
Since we know the number of descendents with $|q,h\rangle_L=|0,2\rangle_L$ of
the chiral primary $|q,q/2\rangle_L$ (namely $4,8,5,2,1$ for $q=0,1,2,3,4$
respectively) we find that the descendants of the chiral primaries
contribute $15397N-54521$ to the $67131N-244053$ states of the
form $ |0,2\rangle_L \otimes |q',q'/2\rangle_R$. Therefore, there are many
states which are not descendants of chiral primaries but 
which do survive the marginal perturbation of the orbifold. 
The first such states are of the form $ |0,1\rangle_L \otimes |q',q'/2\rangle_R$.
The elliptic genus shows that there are $464N-592$ such states,
but only $233N-319$ are descendants of chiral primaries. How to 
account for the missing states in the supergravity description was
the puzzle raised in \cite{vafap}. We will explicitly identify the missing states
of type  $ |0,1\rangle_L \otimes |q',q'/2\rangle_R$ in section~6.2.

\section{The CFT Spectrum versus the KK Spectrum: $(n_L,n_R)=(2,0)$}

\subsection{Single particle states}

The set of $SO(4)$ representations that appears in the compactification
of $(n_L,n_R)=(2,0)$ supergravity on $S^3 \times AdS_3$ was already
computed in (\ref{harm11}). We can compare this to (\ref{decomp}) and find
that the two precisely agree when $t_2=1$, $t_1=0$ and $t_0=n_T+1$.
Thus, the set of short multiplets is of the form
\be \label{e2}
\oplus_m \left(  ({\bf m},{\bf m\pm 2})_S +
(n_T+1) ({\bf m},{\bf m})_S  \right)
\ee   
Since each short multiplet contains precisely one chiral
primary, we find that there are $n_T+1$ chiral primaries of
degree $(m,m)$, one of degree $(m,m+2)$ and one of degree
$(m+2,m)$. These are all single particle states from the point
of supergravity. To construct an arbitrary state in the boundary
conformal field theory we should consider arbitrary products
of these single particle states. Therefore, the complete set
of chiral primaries is obtained by taking arbitrary products
of the single particle chiral primaries. A convenient way to
write down the spectrum is to associate a bosonic creation
operator to each single particle chiral primary and then
to look at the Fock space they create. For each state in
the Fock space there is exactly one chiral primary. This is precisely
the same picture (for $n_T=21$) as we found in conformal field theory
below equation (\ref{e56}). Thus, it seems that there is perfect
agreement as far as the chiral primaries are concerned.

This analysis is, however, restricted to the higher harmonics,
as (\ref{harm11}) is only valid for sufficiently large $m$. 
For example, the $SO(3)$ representation ${\bf 5}$ is only
contained in $({\bf m},{\bf m})$ if $m\geq 3$. The multiplicities
of ({\bf m},{\bf n}) in (\ref{harm11}) are correct
for $m+n>4$, for $m+n \leq 4$ they are given by
\be
\begin{array}{|c|c|c|c|c|c|c|} \hline
\mbox{{\rm $SO(4)$ representation}} & ({\bf 1},{\bf 1}) 
& ({\bf 1},{\bf 2}) & ({\bf 1},{\bf 3}) & 
({\bf 2},{\bf 1}) & ({\bf 2},{\bf 2}) & ({\bf 3},{\bf 1}) \\ \hline
\mbox{{\rm multiplicity }} & 
5n_T+1 & 4n_T+4 & n_T+6 & 4n_T+4 & 6n_T+7 & n_T+6 \\
\hline \end{array}
\ee
If we carefully take these multiplicities into account and
redo the decomposition into short multiplets, we find
that there is one chiral primary of degree $(m,m+2)$ and
one of degree $(m+2,m)$ for $m>0$ only, that there are
$n_T$ chiral primaries of degree $(1,1)$ and $n_T+1$
of degree $(m,m)$ for $m>1$. This is exactly the same as
what we found below (\ref{e56}), except that the chiral
primaries of degrees $(2,0)$ and degrees $(0,2)$ are absent.
This is perhaps not too surprising, as these chiral primaries
correspond in conformal field theory to the 
descendants $J^+_{-1}|0\rangle$
and $\bar{J}^+_{-1}|0\rangle$ of the identity operator.

The way to account for these states is similar to what happens
in four dimensions, where $N=4$ super Yang-Mills is conjectured
to be dual to type IIB supergravity on $S^5 \times AdS_5$. In
\cite{kk1} it was shown that there are $AdS_5$ degrees of freedom that
are pure gauge in the bulk and can be gauged away completely
except at the boundary. These degrees of freedom form a so-called
singleton representation of the relevant AdS supergroup $SU(2,2|4)$, and
the field content of the singleton representation is
precisely that of an $U(1)$ $N=4$ super Yang-Mills multiplet.
This $U(1)$ is naturally identified with the decoupled $U(1)$
of the $U(N)$ $N=4$ super Yang-Mills theory. 

The group $SU(1,1|2)$ does not have singleton representations,
but nevertheless something similar happens in our case. The
$SU(1,1|2) \times SU(1,1|2)$ Chern-Simons theory does not
have any propagating degrees of freedom in $2+1$ dimensions.
However, as discussed in section~3.2, the gauge field is subject
to a boundary condition that contains fields living at the
boundary. These fields are the generators of the left and
right-moving chiral algebra, and their positive frequency modes
make up the representations $({\bf 1},{\bf 3})_S$ and
$({\bf 3},{\bf 1})_S$ of $SU(1,1|2) \times SU(1,1|2)$.
Thus, although these are not singleton representations, they
do correspond to pure gauge degrees of freedom in the bulk.
Including them in the list of short multiplets obtained
from supergravity provides us with the missing chiral 
primaries of degrees $(2,0)$ and $(0,2)$. We now have
a complete and detailed agreement between the chiral
primaries in the orbifold conformal field theory and
those obtained from supergravity.

Although we have no rigorous argument why all KK states should fall
into short representations, it is impossible organize the KK spectrum
differently. It would be interesting to have
a more fundamental understanding of this fact.

\subsection{Multi-particle states}

At this stage one might argue that since the multiparticle states
contain all chiral primaries of the orbifold conformal field theory,
and we also have all generators of the $N=4$ algebra at our disposal,
the complete KK spectrum is exactly equivalent to the set
of chiral primaries and their descendants. If true, this would
lead to a discrepancy between the states obtained from supergravity
and the states that should be present in conformal field theory
according to the elliptic genus calculation of section~5. This puzzle,
raised in \cite{vafap}, can have several solutions, each of
which are somewhat problematic.

(i) There could be a phase transition as one deforms the conformal field
theory from the orbifold point to the actual conformal field theory which
is living at the boundary. The latter has been argued to be a strongly
coupled conformal field theory at zero world-sheet theta angle 
\cite{wit3}, so a priori nothing prevents such a phase transition point
at which the elliptic genus would jump. However, it would be quite awkward to
have so many states appear/disappear at the singularity and 
it is not clear how that could be compatible with unitarity.
Such a drastic jump is 
also not something that occurs when we consider type IIA
on a singular K3 that gives rise to enhanced gauge symmetry.
The corresponding conformal field theory is also strongly coupled,
but we expect only some additional nonperturbative
massless states to appear in the spectrum \cite{aspinwall}.

(ii) Since states with arbitrary large $\bar{L}_0$ contribute to
each term in the elliptic genus, it could potentially receive contributions
from stringy states. This would require stringy states whose spin $s$
and mass $m$ satisfy $s \sim m \ell \sqrt{\alpha'}$. It is hard to
imagine such states with an enormous spin arising from string theory.

(iii) All the additional states could live purely on the boundary of
$AdS_3$ and therefore correspond to ``singletons'', in the same
way as the stress-tensor of the boundary theory corresponds to
a singleton. Singletons correspond to pure gauge degrees of freedom
in the bulk, and there does not seem to be room for such a large
number of additional gauge fields in six-dimensional supergravity
on $S^3 \times AdS_3$. Alternatively, the duality involving supergravity
could be incomplete and the proper duality would require us to add the
additional degrees of freedom by hand. In that case, there would be
something crucial missing from the otherwise quite successful solitonic
description of brane configurations. 

From all these points of view it would be much nicer if all states would
already be present in the supergravity description and we will now
argue that in fact they are. 

The main point is the fact that although the product of two chiral
primaries is again a chiral primary, this is not true for their
operator product expansion. The leading term is a chiral primary,
but the subleading regular terms contain non-chiral primaries as
well. One easy way to see this is by looking at the transformation
properties of characters of $N=4$ representations under modular
transformations \cite{n4char}. An equivalent way to put this is
that correlation functions of chiral
primaries contain a sum over all fields in the intermediate
channels, not just over the chiral primary ones. Now
according to the proposal of \cite{wit,gkp}, we can compute
arbitrary correlators in the boundary theory by taking suitable
boundary conditions for the fields in the bulk and
computing the partition function for the bulk theory. 
It is then straightforward to compute correlators of non-chiral
primaries: one simply considers correlators of chiral
primaries, lets the arguments of the chiral primaries approach
each other and subtracts out the leading singularities from
the correlation function. For example, suppose the operator product
expansion of the chiral primaries $A$ and $B$ contains the
chiral primary $C$ and the non-chiral primary $D$ as
\be \label{exa}
A(z) B(w) \sim C(w) + (z-w) D(w) + \ldots
\ee
The two-point function of $D$ can then be computed as
\be
\langle D(x) D(y) \rangle = 
\lim_{z\rightarrow x} \lim_{w \rightarrow y}
\left\langle
\frac{A(z) B(x) - C(x)}{z-x}
\frac{A(w) B(y) - C(y)}{w-y} \right\rangle
\ee
The right hand side can be obtained directly from the prescription
of \cite{wit,gkp}.
Although this may seem somewhat indirect, all information about
the conformal field theory is in this way present in the
supergravity description. 

There is a different way to state the above which is closer to
the supergravity description. Recall that in the supergravity
description the various single particle fields were organized
in representations of $SU(1,1|2) \times SU(1,1|2)$. Multiparticle
states are products of single particle states, transforming
in the tensor product representation of $SU(1,1|2) \times SU(1,1|2)$.
The tensor product of two short representations of
$SU(1,1|2)$ does not contain just a single short representation
of $SU(1,1|2)$, but various longer ones as well. These longer
ones correspond to the non-chiral primaries. In other words,
the product of two descendants of chiral primaries is not
necessarily the descendant of a chiral primary. Indeed, in
(\ref{exa}) the field $D(z)$ is nothing but $(\partial A(z)) B(z)$,
and we could also have computed the $D$-two point function via
\be
\langle D(x) D(y) \rangle =
\lim_{z\rightarrow x} \lim_{w \rightarrow y}
\left\langle
\partial A(z) B(x) 
\partial A(w) B(y) \right\rangle
\ee  

As further evidence that this is the correct interpretation of
multi-particle states, we will show in an example that this
procedure accounts for all additional states predicted by
the elliptic genus. Consider states of the form 
$|0,1\rangle_L \otimes |q',q'/2\rangle_R$.
The elliptic genus shows that there are $464N-592$ such states,
but only $233N-319$ are descendants of chiral primaries, as
we mentioned at the end of section~5. The only way to
make states with $q=0$ and $h=1$ as multiparticle states
is to take the product of two states with $h=1/2$, which
can only appear as descendants of chiral primaries with
$q=1$ and $q'$ arbitrary. The cohomology classes of
weight $(1,q')$ are: (i) for each $i$, $0\leq i \leq N-1$,
there are 20 forms of degree $(1,1) \otimes (0,2)^i 
\equiv (1,1+2 i)$ ($(0,2)$ represents the anti-holomorphic
two-form), (ii) for each $i$, $0 \leq i \leq N-2$, there
is one form of degree $(1,1)\otimes (0,2)^i\equiv(1,1+2i)$
and one of degree $(1,3) \otimes (0,2)^i \equiv (1,3+2 i)$.
The total number of forms 
of degree $(1,q')$ 
is $20N+2(N-1)=22N-2$, which
is responsible for the appearance of this factor in
(\ref{ellexp}) and (\ref{coho}). Each of the $SU(1,1|2)$
representations corresponding to a form of degree $(1,q')$
contains a spin-$1/2$ doublet of states with $h=1/2$,
namely $|1,1/2\rangle_L \otimes |q',q'/2\rangle_R$
and $|-1,1/2\rangle_L \otimes |q',q'/2\rangle_R$.
The tensor product of two of such spin-1/2 representations
contains a spin-1 representation with descendants of
a chiral primary, and a singlet, which is a non-chiral
primary of weight $|0,1\rangle_L \otimes |q',q'/2\rangle_R$.

To count the number of non-chiral primaries, denote the
forms under (i) above by $\beta_k \bar{\Omega}^i$, with
$\beta_k$, $k=1,\ldots,20$ the 20 $(1,1)$ forms and $\bar{\Omega}$
the anti-holomorphic two-form. The forms under (ii) will be
denoted by $\alpha_l \bar{\Omega}^i$, $l=1,2$. The forms
$\beta_k$ originate from forms on $K3$ and can therefore
be multiplied by at most the $(N-1)$th power of $\bar{\Omega}$,
whereas $\alpha_l$ originate from forms on $K3^2/S_2$ and
can therefore be multiplied by at most the $(N-2)$th power
of $\bar{\Omega}$. From the product of $\beta_k \beta_l \bar{\Omega}^i$
we get one non-chiral primary for each $k>l$, and since $\beta_k \beta_l$
should be thought of as living on $K3^2/S_2$, it can be
multiplied by at most the $(N-2)$th power of $\bar{\Omega}$.
Therefore, the total number of non-chiral primaries obtained
this way is $\frac{19 \cdot 20}{2}(N-1)$. Similarly, the forms
$\beta_k \alpha_l$ live on $K3^3/S_3$ and give 
total of $20 \cdot 2 \cdot (N-2)$ 
non-chiral primaries. Finally, $\alpha_1 \alpha_2$ lives on
$K3^4/S_4$ and gives upon multiplication with $\bar{\Omega}^i$ a
total of $N-3$ non-chiral primaries. Altogether the 
total number of non-chiral primaries of the form
$|0,1\rangle_L \otimes |q',q'/2\rangle_R$ obtained this way is
\be
190(N-1) + 40(N-2) + (N-3) = 231N-273
\ee
Together with the $233N-319$ descendants of chiral primaries we
find a total number of $464N-592$ states of the form 
$|0,1\rangle_L \otimes |q',q'/2\rangle_R$, which is precisely the
number predicted by the elliptic genus. We consider this as
strong evidence in favor of our proposal.

\section{Other 6d and 5d Supergravities and their KK Spectrum}

\subsection{6d supergravity with $n_L+n_R>1$}

In this section we briefly redo the analysis for the other 6d
supergravities compactified on $S^3 \times AdS_3$. In all
cases we find that the KK spectrum agrees with that of a
conformal field theory whose target space is the symmetric
product of some four-manifold. Furthermore, the KK spectrum
allows us to read of the cohomology of the four-manifold. 
One new feature is the appearance of short multiplets $({\bf m},{\bf m'})_S$
with $m+m'$ odd. The chiral primaries in these short multiplets come
from fermionic fields in the supergravity, and correspond to odd-degree
forms in the target space of the non-linear sigma model. From
either point of view it is clear that we should associate 
fermionic creation rather than bosonic creation operators
to these chiral primaries, in complete agreement with 
(\ref{Poincare}) and the discussion below (\ref{e56}). 
A second new feature is the appearance of
additional singletons, related to the additional 
supersymmetries. The singletons give rise to additional
short multiplets of the form $({\bf 1},{\bf 2})_S$ and
$({\bf 2},{\bf 1})_S$ on the boundary. 

The general result is
that if the KK spectrum, when decomposed into short multiplets,
is for sufficiently high harmonics
given by (\ref{e1})
\be
\oplus_m ( 
t_2 ({\bf m},{\bf m \pm 2}) 
+ t_1 ({\bf m},{\bf m \pm 1})_S
 + 
t_0 ({\bf m},{\bf m})_S 
)
\ee
then the hodge diamond of the four-manifold looks like
\be \label{hodge}
\begin{array}{ccccc}
 & & 1 & & \\
 & t_1/2 & & t_1/2 & \\
 t_2 & & t_0-t_2-1 & & t_2 \\
 & t_1/2 & & t_1/2 & \\
 & & t_2 & & 
\end{array}
\ee
In certain cases the Hodge diamond will not make any sense, but only in
cases in which the corresponding supergravity does not exist, so this
is not something to worry about. We now give the
values of $t_0,t_1,t_2$ for the various supergravities.

$(n_L,n_R)=(2,2)$. This is the case with the maximal amount of
supersymmetry, which is obtained for example by compactifying
type IIA or IIB supergravity on $T^4$.
The field content consists of a graviton, eight
gravitino's, 5 self-dual and 5 anti self-dual two forms, 16 gauge
fields, 40 fermions and 25 scalars. 
The KK spectrum of $SO(4)$ representations 
is found to be
\be
\oplus_m \left( ({\bf m},{\bf m\pm 4}) + 8({\bf m},{\bf m\pm 3}) + 28({\bf m},{\bf m \pm 2})+
 56(n_T+2) ({\bf m},{\bf m \pm 1}) + 70({\bf m},{\bf m}) \right) .
\ee
After organizing this in terms of short multiplets we find
$t_0=6$, $t_1=4$ and $t_2=1$. The hodge diamond (\ref{hodge})
is that of $T^4$, showing that the conformal field theory is
a sigma model with target space a symmetric product of $T^4$,
as expected. This theory was recently studied in \cite{mart},
and is an example where we have fermionic short multiplets 
and additional singletons.

$(n_L,n_R)=(1,1)$. 
 This theory, with $n_V$ vector multiplets and
$n_U$ $USp(2)$ vector multiplets (whose field content is given
in \cite{stra}) yields $t_0=n_V + 3 n_U + 2$, $t_1=0$ and
$t_2=1$.  An example of such a
theory is obtained by putting type IIA on K3, which is a theory with
$20$ vector multiples and $n_U=0$. We find that the cohomology is
precisely that of $K3$. Thus, type IIA supergravity 
on $K3\times S^3 \times AdS_3$
seems to be dual to type IIB on a different $K3\times S^3 \times AdS_3$.

$(n_L,n_R)=(3,0)$. 
Two multiplets are listed in \cite{stra}. Taking
$n_1$ of the first and $n_2$ of the second yields
$t_1=n_1 + 2n_2$, $t_2=n_2$ and
$t_0=2(n_1+n_2)$. 

$(n_L,n_R)=(2,1)$. Except for the gravity multiplet the theory has one
more multiplet that includes a spin-$3/2$ state. Taking $n$ of these
multiplets, we get $t_1=n+2$, $t_2=1$ and
$t_0=2n+2$.         

$(n_L,n_R)=(3,1)$ or $(4,0)$.
These supersymmetry algebras have only a single
acceptable representation, of which we take $n$. The cohomologies come
out as
(\ref{hodge}) with
 $t_1=4 n$, $t_2=n$ and $t_0=6n$.

\subsection{6d supergravity with $n_L+n_R=1$}

The most interesting case is the 6d supergravity with the smallest number
of supersymmetries, i.e. the theory with 
 $(n_L,n_R)=(1,0)$, on $S^3 \times AdS^3$.
Such six-dimensional supergravities are obtained either from
F-theory on a Calabi-Yau manifold, or from heterotic or 
type I string theory on $K3$. Although we have not verified this,
this six-dimensional supergravity on $S^3 \times AdS_3$ could
for instance describe 
the near-horizon geometry of
a $D1$-$D5$ system in type I theory on
$K3$, with the $D5$ branes wrapping the $K3$. On the $D1$ branes
we expect to find a sigma model with $(4,0)$ supersymmetry
of the type discussed in \cite{wit7,doug1}. Sigma models with
$(4,0)$ supersymmetry are similar to those with $(4,4)$ 
supersymmetry, except that the right moving fermions
live in some vector bundle which is not the tangent bundle,
and couple to a self-dual gauge field for this vector bundle
in the world-sheet lagrangian. In our case we expect to
get a sigma model whose target space is the moduli space
of $Sp(N)$ instantons on $K3$, with the right-moving fermions
coupling to some vector bundle over this moduli space. The
spectrum of this theory does not only involve the cohomology
of the instanton moduli space, but also various vector bundle
cohomologies as in \cite{disgr}. We do not know explicit results
for these cohomologies, making it difficult to determine the
CFT spectrum. What is even more problematic is the lack of a
precise definition of a chiral primary.
The left-movers have an $N=4$
superconformal algebra, and we can certainly take the
usual definition of a chiral primary for the left-movers,
but the right-movers have only an $SU(2)$ current algebra
and a Virasoro algebra. Nevertheless we can go ahead
and compute the KK spectrum of the supergravity theory,
which will give us a prediction for the spectrum of the
conformal field theory. Using  the techniques used so
far, we can no longer determine the $\bar{L}_0$-eigenvalue
of the various KK-fields. However, the spin of fields on
$AdS_3$ is equal to the difference of the $L_0$ and
${\bar L}_0$ eigenvalue, and the spins can be determined as
follows. Given a field transforming in some representation
$({\bf n_1},{\bf n_2})$ of the little group $SO(4)$,
we decomposed it in $SO(3) \subset SO(4)$ representations,
and subsequently found the $SO(4)$ representations
$({\bf m},{\bf m+d})$ that yield the same $SO(3)$
representations for $SO(3) \subset SO(4)$. Notice that
these two $SO(4)$'s are different; one is the little
group in six dimensions, the other one is the isometry group
of the sphere. The representation $({\bf n_1},{\bf n_2})$
contains various $U(1) \times U(1) \subset
SU(2) \times SU(2)=SO(4)$ representations, with
$U(1) \times U(1)$ eigenvalues $y_1,y_2$. The
$U(1)$'s are normalized so that the eigenvalues
are half-integer. Then the spins of the various
fields associated to $({\bf m},{\bf m+d})$ are
the possible values of $y_1-y_2$ given that 
$y_1+y_2=d$. Incorporating these spins in
the determination of the short multiplets enables
us to determine the various $\bar{L}_0$ 
eigenvalues.

Six dimensional $(1,0)$ supergravity theory
can in general have $n_H$ hyper multiplets, $n_V$
vector multiplets and $n_T$ tensor multiplets. 
For a discussion of the equations of motion, see \cite{eom}.
Anomaly cancellation
implies $29 n_T + n_H - n_V = 273$, but this will not be relevant
as we are only interested in classical supergravity. 
We will assume that the gauge group is a product of $U(1)$'s,
with respect to which the hypermultiplets are neutral.
When we compactify the theory
on $S^3 \times AdS^3$, the KK spectrum of $SO(4)$ representations
reads for sufficiently large $m$
\bea \label{aa}
& & \oplus_m \left(
({\bf m},{\bf m\pm 4}) + 2 ({\bf m},{\bf m\pm 3}) + 
(n_T+n_V+3) ({\bf m},{\bf m\pm 2}) \right. \nonumber \\
& & \qquad \left. + (2(n_T+n_H+n_V)+4) ({\bf m},{\bf m\pm 1}) + 
 2(n_T + 2n_H + n_V + 2)({\bf m},{\bf m}) \right).
\eea
We want to organize this in representations of 
$SU(1,1|2) \times Sl(2,{\bf R}) \times SU(2)$. 
We will denote by $({\bf m},{\bf m'};s)_S$ the tensor product
of the short representation (\ref{short}) with $k=m-1$
of $SU(1,1|2)$, the ${\bf m'}$-dimensional representation
of $SU(2)$ and a
highest weight representation of $Sl(2,{\bf R})$
with highest weight $\frac{m-1}{2}-s$. 
Thus $s$ is the difference between the conformal
weight of the left-moving chiral primary and the
right-moving primary.
Decomposing (\ref{aa}) in these short representations yields
\bea \label{pq1}
& & \oplus_m \left(
({\bf m},{\bf m+2};-1)_S  
+ n_T ({\bf m},{\bf m};0)_S
+ n_V ({\bf m},{\bf m};-1)_S
\right. \nonumber \\ & & \qquad \left.
+ ({\bf m},{\bf m};0)_S
+ ({\bf m},{\bf m};-2)_S
+ 2 n_H ({\bf m},{\bf m-1};-1/2)_S
\right. \nonumber \\ & & \qquad \left.
+ n_T ({\bf m},{\bf m-2};-1)_S
+ n_V ({\bf m},{\bf m-2};0)_S
+ ({\bf m},{\bf m-2};-1)_S
\right. \nonumber \\ & & \qquad \left.
+ ({\bf m},{\bf m-2};1)_S
+ ({\bf m},{\bf m-4};0)_S \right)
\eea

This result is not very transparent as it stands, but becomes
very suggestive when we consider six-dimensional supergravity
obtained from F-theory on a Calabi-Yau threefold $M$, and
forget the spin dependence in (\ref{pq1}). 
The number of multiplets is
expressed in terms of the cohomology of $M$
and the cohomology of the base $B$ over which
$M$ is elliptically fibered,
namely  $n_T=h^{1,1}(B)-1$, $n_V=h^{1,1}(M)-h^{1,1}(B)-1$,
and $n_H=h^{2,1}(M)+1$ \cite{fth}, if we assume the gauge group
is a product of $U(1)$'s and all matter is neutral.
The representations (\ref{pq1}) can very simply be written
in terms of the hodge numbers $h^{p,q}(\tilde{M})$ of the
mirror Calabi-Yau
\be \label{f04}
\oplus_m\oplus_{i=-3}^3  \left( 
 (\sum_k h^{k+i,k}(\tilde{M}) ) ({\bf m},{\bf m+i-1}) 
\right)
\ee
In fact, this result turns out also to be correct in case
$M=K3\times T^2$ and $M=T^6$. It seems that the
$(0,4)$ conformal field theory knows about the duality 
between type I and F-theory, and one can 
associate in a natural way a $(0,4)$ conformal field theory
to any Calabi-Yau. It would be interesting to understand
this at a more fundamental level, and whether in other cases
of string-duality the conformal field theory of the dual
theory can be constructed by putting the original
theory on a certain $AdS$ space. It would also be
interesting to examine the $(0,4)$ theories themselves in some
more detail, perhaps along the lines of 
\cite{dgm}.

\subsection{5d supergravity on $S^2 \times AdS_3$}

According to one of the conjectures of \cite{mal}, M-theory
compactified on $M\times S^2 \times AdS_3$ with $M$ some
Calabi-Yau manifold is dual to a $(0,4)$ superconformal
field theory on the boundary of $AdS_3$. The central charge
of the $(0,4)$ theory can be used to compute the entropy of
the corresponding 4d black hole \cite{bh5,vaf5}. The $(0,4)$
theory lives on an M-theory fivebrane wrapping some holomorphic
four-cycle $C$ in $M$, and the left-movers have central charge
$b^{\rm even}(C)+b^{\rm odd}(C)$. 

In order to see to what extent M-theory on $M\times S^2 \times
AdS_3$ knows about this, we compute its KK spectrum using
the techniques explained in section~4 and section~7.2,
 but now applied to
five dimensions. The anti-de Sitter supergroup is
$SU(1,1|2) \times Sl(2,{\bf R})$, and by $({\bf m};s)_S$ 
we denote the tensor product of a short representation
(\ref{short}) with $k=m-1$
of $SU(1,1|2)$ and a
highest weight representation of $Sl(2,{\bf R})$
with highest weight $\frac{m-1}{2}-s$. As before,
$s$ is the difference between the conformal weight
of the left-moving chiral primary and the right-moving
primary. 
For five-dimensional $N=1$ supergravity with $n_H$ hypermultiplets
and $n_V$ vectormultiplets compactified on $S^2 \times AdS_3$
we get for the KK spectrum (ignoring singletons)
\bea \label{ab1}
& & 
n_H ({\bf 2};-1/2)_S + 
n_V ({\bf 3};-1)_S +
n_V ({\bf 3};0)_S +
({\bf 3};-2)_S +
({\bf 3};1)_S \nonumber \\ & & 
+ \oplus_{m>1} (n_H ({\bf 2m};-1/2)_S +
n_V ({\bf 2m+1};-1)_S +
n_V ({\bf 2m+1};0)_S +
({\bf 2m+1};-2 )_S \nonumber \\ & & \qquad \qquad
+ ({\bf 2m+1};-1 )_S
+ ({\bf 2m+1};0 )_S
+ ({\bf 2m+1};1 )_S ). 
\eea
The number of vector multiplets equals $h^{1,1}(M)-1$, and
the number of hyper multiplets is $2(h^{1,2}(M)+1)$ \cite{5dsugra}.
Thus, dropping the spin dependence (\ref{ab1}) can be rewritten as 
\be \label{ab2}
b^{\rm odd}(M) ({\bf 2})_S + (b^{\rm even}(M)-2) ({\bf 3})_S +
\oplus_{m>1} (b^{\rm odd}(M) ({\bf 2m})_S + b^{\rm even}(M) ({\bf 2m+1})_S ).
\ee
The KK spectrum (\ref{ab1}) and its simplified
version (\ref{ab2})  do not depend on the cohomology of the holomorphic four-cycle,
but only on the cohomology of the Calabi-Yau manifold $M$. The information about 
the four-cycle is therefore not encoded in the KK spectrum of supergravity, but should
manifest itself in the interactions and correlation functions
of the conformal field theory. For instance, the central charge can be
read off from the two-point function of the stress-energy 
tensor\footnote{For a recent discussion of the central charge, see
\cite{cen1,cen2}.}. 
It would be interesting to understand this in more detail.

\section{Conclusions}

In this paper we have shown that there is detailed agreement
between the Kaluza-Klein spectrum of supergravity and
the spectrum of certain conformal field theories. To account
for all the states in conformal field theory we had to consider
products of descendants of chiral primaries. It is an interesting
question what the role of the analog fields in four-dimensional
$N=4$ $d=4$ super Yang-Mills theory is. The relevant supergroup
in four dimensions is $SU(2,2|4)$, and there are many fields
that can be constructed as products of descendants of chiral
primaries. If their dimensions are still protected, we can perhaps
learn something about 
the singularity structure of the 4d
correlation functions.

In two dimensions there are many things that deserve a better
understanding. In particular,
the cases with 8 supersymmetries in section~7.2 and~7.3
are still somewhat mysterious. We would also like to
know the implications of our results for the entropy 
and various emission and absorption probabilities
of 4d and 5d black holes. An issue we have not discussed
is the RR sector of the conformal field theories.
To study those the $AdS_3$ part has to be replaced by
a certain three-dimensional black hole \cite{uuh}.
It would be interesting to redo the KK analysis in
that background to have an independent test of the
conjectured duality. Finally, the techniques can
easily be generalized to study other theories
involving $AdS_3$ or other $AdS$ spaces. 
In the examples studied in this paper, the KK
spectrum turned out to be almost completely
determined by the field content and symmetries
of the theory. If true in general, this would
imply that KK spectra are not a very deep test
of the $AdS\leftrightarrow$CFT conjecture,
and that most of the interesting information
is hidden in the interactions of the theory.

\noindent
{\bf Acknowledgement}

I would like to thank O. Aharony, P. Berglund, S. Ferrara, 
J. Gomis, K. Hori, C. Johnson, S. Kachru, A. Karch, M. Li, E. Martinec,
P. Mayr, Y. Oz, S. Shenker, K. Skenderis, P. Townsend, C. Vafa and
especially H. Ooguri for discussions,
and the ITP at Santa Barbara
for hospitality. 
This work is supported in part by NSF grant PHY-9514797 and DOE
grant DE-AC03-76SF00098.
The author is a fellow of the Miller Institute for Basic Research
in Science.

\end{document}